# What's (Not) Validating Network Paths: A Survey

KAI BU and YUTIAN YANG, Zhejiang University
AVERY LAIRD, Simon Fraser University
JIAQING LUO, University of Electronic Science and Technology of China
YINGJIU LI, Singapore Management University
KUI REN, Zhejiang University

Validating network paths taken by packets is critical for a secure Internet architecture. Any feasible solution must both enforce packet forwarding along endhost-specified paths and verify whether packets have taken those paths. However, neither enforcement nor verification is supported by the current Internet. Due likely to a long-standing confusion between routing and forwarding, only limited solutions for path validation exist in the literature. This survey article aims to reinvigorate research in to the significant and essential topic of path validation. It crystallizes not only how path validation works but also where seemingly qualified solutions fall short. The analyses explore future research directions in path validation toward improving security, privacy, and efficiency.

## 1 INTRODUCTION

> "To remember where you come from is part of where you're going."
> — *Anthony Burgess*

Validating network paths is indispensable for a secure Internet architecture [5, 11, 29]. Unlike the current Internet, where routers have full control over packet delivery, path validation empowers end hosts with more control. It enables end hosts to enforce the paths they would prefer their packets to follow. End hosts can also verify whether these paths are really followed. Clearly, any

This work is supported by the National Science Foundation of China under Grant 61402404 and Grant 61602093.
Author's addresses: K. Bu and Y. Yang, College of Computer Science and Technology, Zhejiang University, 38 Zheda Road, Hangzhou 310027, China; email: {kaibu, ytyang}@zju.edu.cn; A. Laird, School of Computing Science, Simon Fraser University, Burnaby V5A 1S6, Canada; email: alaird@sfu.ca; J. Luo, School of Computer Science and Engineering, University of Electronic Science and Technology of China, Chengdu 611731, China; email: csjluo@hotmail.com; Y. Li, School of Information Systems, Singapore Management University, 80 Stamford Road, Singapore 178902, Singapore; email: yjli@smu.edu.sg; K. Ren, Zhejiang University, 38 Zheda Road, Hangzhou 310027, China; email: kuiren@zju.edu.cn.

feasible path validation solution would modify packet processing logic on routers. This makes path validation still transcendental for future Internet architectures [11]. Then one may wonder: Why bother with the so caused remanufacture efforts? We can already enjoy diversified high quality Internet services anyway. The transparency of underlying packet delivery, however, necessitates path validation. Imagine a supposed life-saving call to 911 that gets answered a bit late or even fails to get through. Should the unfortunate caller attribute his or her unluckiness to the slow Internet backbone [14] for not processing his or her call more quickly? Or should the caller question whether the call is directed along a slower path that is not intended for critical services like 911? This concern is confirmed a decade ago—Internet Service Providers (ISPs) may sometimes transit each other's traffic for profits [91]. In particular, the ISP you signed up with may direct your traffic to other ISPs with inferior performance. However, the current Internet provides no means for users to enforce and verify packet delivery. Such a deficiency might be exploited for security breaches over, for example, financial, medical, and military services built on the Internet [9]. To mitigate this security issue, the Internet community advocates validating network paths by enforcing where packets should visit and verifying where packets have transited. With path validation, we expect to make packet delivery more reliable and robust.

Path validation is not the first to tackle the insecurity of the current Internet, but it is the most robust. The first to rally against an insecure Internet is secure routing [44]. It aims to find the best (e.g., shortest) paths connecting end hosts without the path finding process being attacked. However, finding a best path does not necessarily mean that packet delivery strictly follows the best path. Under the control of attackers, compromised routers may discard or detour packets. To improve forwarding reliability, source routing embeds path directives in packet headers [35]. A path directive is a sequence of routers that should be visited, in order, by the corresponding packet. Traceroute [68] enables end hosts to retrieve the path taken by packets. Only with sufficient cryptography augmentation, secure source routing [9] and secure traceroute [60] can enforce and verify network paths, respectively, without being vulnerable to forging and tampering. Both solutions, however, are not designed with each other taken into account. This makes them hard to operate concurrently. It is also infeasible or inefficient to simply combine them for path validation.

This paper presents the first comprehensive survey of path validation. Simply put, path validation functions by embedding cryptographic states computed over path metadata in packet headers [58]. Such packet-carried states are initialized by the source and force enroute routers to forward packets along the source-specified path. Moreover, they are updated on enrouter routers, enabling downstream routers and the destination to verify path compliance. Clearly, path validation increases packet size and thus imposes bandwidth overhead on a network. This overhead, however, is deemed insignificant to plentiful network bandwidth [20]. The benefits in performance, reliability, and security we gain from path validation far outweigh the additional bandwidth overhead. In spite of the significance to advancing Internet technology, path validation has a surprisingly limited number of solutions. We ascribe this rarity to a lack of systematic distinction among Internet enhancement solutions: almost all of them claim reliable end-to-end communication as a goal. Without a deep understanding of each solution, we may easily miss subtle yet essential differences that render a solution unsuitable for path validation. This further motivates us to conduct this survey. When identifying path validation solutions, we adhere to the strict requirement that path validation should jointly achieve path enforcement and path verification [46]. Path validation design evolves toward improving efficiency with reduced state size. Sporadic work [87] is dedicated to analyze the security properties using formal security proofs. A variant also validates where packets should not transit [48]. Inspired by the disparity of established efforts for path validation and the attention it should deserve, we also explore future research avenues toward enhancing efficiency, security, and privacy.

It is challenging to conduct such a thorough and systematic survey, particularly when filtering unqualified solutions and scrutinizing qualified ones. As aforementioned, many solutions may state their design goals similar to that of path validation, only to fall short. Often, they either refer to secure routing instead of forwarding [44], or partially solve path validation with only one attempt of path enforcement [9] and path verification [60]. These unqualified solutions far outnumber the qualified ones, making the classification process nebulous and difficult. Even for path validation solutions that jointly enforce and verify network paths, we have to scrutinize certain subtle differences therein. We find that later solutions do not gain efficiency for free, but rather at the cost of security. This, however, can hardly be noticed without a systematic understanding of path validation solutions.

We strive for crystallizing what works for path validation, what does not, and what is worth pursuing. Table 1 classifies all solutions to be reviewed in this paper. We highlight our major contributions as follows, with sincere hope that they may benefit academic researchers and industry professionals with useful directions for validating network paths and therefore securing the Internet.

- Establish the motivation and requirements for path validation. This helps readers to better understand the significance of path validation.
- Clarify the differences between path validation solutions and various other Internet enhancements such as secure routing, (secure) source routing, and (secure) traceroute. Only after filtering unqualified solutions can we concentrate on how path validation works, evolves, and limits.
- Scrutinize the principles and limitations of path validation solutions. We find that their efficiency gains come at the expense of security. This subtle caveat can only be uncovered thorough a systematic understanding of path validation solutions.
- Explore future research avenues and suggest feasible directions. Path validation has many breakthroughs in efficiency, security, and privacy to come.

The rest of the paper is organized as follows. Section 2 motivates the path validation problem and specifies solution requirements. Section 3 covers related work that is concerned with loose control over network traffic but fails to validate their forwarding paths. Section 4 reviews path validation solutions. Section 5 focuses on a variant of the path validation problem called alibi routing. Section 6 identifies open issues and future directions. Finally, Section 7 concludes the paper.

## 2 PATH VALIDATION: WHAT AND WHY

In this section, we present the motivation and requirements for validating network paths. The current Internet focuses more on the process of finding routes or paths between Autonomous Systems (ASes) [50]. However, the issues of how to enforce packet forwarding along a specified path, and how to verify whether that path was really taken, are less explored. All network entities along a path—the source, intermediate routers, and the destination—have motivations for enforcing and verifying network paths (Section 2.2). When considering potential points of failure ranging from external attackers to misbehaving internal network entities, it becomes clear that path enforcement and verification are both essential prerequisites for a robust Internet. The solutions, referred to as path validation [46], are indispensable for a secure Internet architecture.

### 2.1 Out of Source, Out of Control: Packet Forwarding under Internet's Status Quo

In the current Internet architecture, routers decide how to forward a packet toward its destination [26]. A forwarding decision depends on a router's local routing table. Each entry in the routing

Table 1. Reference classification.

| Category | Reference | Index | Principle |
|---|---|---|---|
| Secure Routing[1] | S-BGP [44] | §3.1.1 | cryptographically secure routing messages against prefix hijacking, route spoofing, and eavesdropping |
| | SPV [39] | | |
| | SCION [88] | | |
| | Lychev et al. [50] | | |
| Source Routing | i3 [72] | §3.1.2 | embed plain-text path directives in packet headers to direct packet delivery on enroute routers |
| | DOA [77] | | |
| | NUTSS [34] | | |
| | Dysco [86] | | |
| | NIRA [83] | | |
| | RBF [62] | | |
| | Multipath Routing [31, 35, 55, 84, 90] | | |
| Traceroute | PPM [68] | §3.1.3 | mark router history in packet headers, or log packet history on routers, to retrieve taken paths |
| | Efficient PPM [33, 71, 81] | | |
| | DPM [22] | | |
| | Cherrypick [76] | | |
| | SPIE [69] | | |
| | NetSight [37] | | |
| | Hybrid Design [25, 51, 52, 82] | | |
| Path Verification: secure traceroute | Padmanabhan et al. [60] | §3.2 | cryptographically secure packet marks or logs against forging & tampering |
| | Wong et al. [79] | | |
| | AudIt [7] | | |
| | RPVM [41] | | |
| | SPP [23] | | |
| Path Enforcement: secure source routing | Platypus [63] | §3.3 | cryptographically secure packet-carried path directives against forging & tampering |
| | Avramopoulos et al. [9] | | |
| | Ethane [21] | | |
| | ARROW [61] | | |
| | Onion Routing [75] | | |
| | Tor [74] | | |
| | HORNET [24] | | |
| Path Validation: enforcement + verification | PFRI [20] | §4.1 | embed cryptographic states in packet headers; states enforce specified paths; routers update states toward verifying path compliance |
| | ICING [58] | §4.2 | |
| | OPT [46] | §4.3 | |
| | OSV [18, 19] | §4.4 | |
| | Alibi Routing [48] | §5 | |
| Notes: | | | |
| 1. We list only typical secure routing solutions. More are reviewed in [3, 17, 38, 40, 50, 59]. | | | |

table associates a reachable destination with a next hop router along the path to the destination. An entry may also contain fields like path cost and quality of service that drive path selection. Routing tables can be configured either manually or dynamically. Manual configuration requires that a network administrator calculate all paths for potential packets across the network. For each path, the administrator needs to add a routing table entry on every router along the path. This task

is manageable for small and static networks. However, it becomes challenging, tedious, and error prone when networks are large or dynamic. Therefore, most routers adopt dynamic configuration that uses routing protocols to automatically configure routing tables and update them upon network changes. Routing protocols enable routers to propagate network topology such that any router may locally determine a path. Routing protocols for an individual AS are referred to as Interior Gateway Protocols (IGPs), while those for communicating different ASes are referred to as Exterior Gateway Protocols (EGPs) [53]. The mainstream routing protocols currently in use are the Open Shortest Path First (OSPF) protocol [56, 57] for IGPs and the Border Gateway Protocol (BGP) [64] for EGPs. The software layer running routing protocols in modern router architectures is called the control plane. It is the hardware layer, called the forwarding plane, that forwards packets in line speed. To this end, routers need to compile routing tables in the control plane into forwarding tables in the forwarding plane. The medium storing forwarding tables should support fast lookup. Common such mediums in use are Static Random Access Memory (SRAM), Dynamic Random Access Memory (DRAM), and Ternary Content Addressable Memory (TCAM). The lookup process usually follows the longest prefix match algorithm, which forwards a packet to the next hop based on whichever routing table entry has the most leading-edge bits in common with the destination address [65].

Once packets depart from the source, it has no control over how enroute routers may forward packets. The destination is just as powerless; neither entity has the ability to enforce path preferences, that is, which path their packets should follow. Moreover, the lack of packet verification on routers makes forwarding exploitable for many attacks. For example, since routers do not authenticate packet sources, IP spoofing based DDoS attacks forge attacking packets' source addresses, which are hard to trace back [68]. Compromised routers can also arbitrarily tamper packet payloads without detection by downstream routers.

While source authenticity and packet integrity can be verified based on end-to-end cryptographic authentication, network paths taken by packets cannot be easily verified or enforced in the current Internet. Verifying paths using either per-hop per-packet receipts to the source (e.g., traceroute [69]) or per-hop per-packet logs for query [23] incurs an impractically heavy overhead of communication or storage. Enforcing paths is even harder: In the current Internet, packets carry an indication of only where they should go (i.e., destination addresses in packet headers) without any packet-specific preferences for intermediate routers. One may suggest that before sending packets, the source or destination informs each enroute router of its preferred path, with packet identifiers and preferred forwarding decisions. Then, upon receiving a packet, each router locally matches it against the recorded packet identifiers and, once a match is found, follows the associated forwarding decision. Again, this solution causes heavy communication and storage overhead to network channels and elements. Besides the efficiency concern, a bigger challenge comes when compromised routers may misbehave [46, 58].

### 2.2 My Packet, My Way: That's when You Can't Count on Current Internet

Before detailing the requirements and challenges for enforcing and verifying network paths, we first explore their use cases. Certain demands must be met for all three types of entities along a network path, that is, the source/sender, intermediate/enroute routers, and the destination/receiver.
**Source's demands.** The source specifies path preferences mainly for guaranteeing quality of service. Expected service quality is usually specified in a Service Level Agreement (SLA) agreed with the service provider. For example, ISPs provide customers with multiple choices of network bandwidth, such as 100 Mbps and 1 Gbps. Similarly, mobile carriers offer different wireless bandwidth to customers through different cellular technologies, such as 3G and 4G. However, a misbehaving

ISP/carrier may forward traffic via an inferior path rather than the premium one promised in the SLA [41, 46]. Moreover, different ISPs may sometimes transit traffic between each other [91]. In either case, a customer may experience degraded service quality due to path noncompliance.

The source may also specify path preferences for additional security and privacy. It is common that certain regions deploy censorship systems to inspect and filter Internet traffic. When the source in a region sends packets to the destination in another region, the packets may transit a series of intermediate regions. If there are multiple feasible paths transiting different sequences of intermediate regions, the source likely wishes to choose one without censorship. This is more challenging to address because we need to prove not only that a packet takes the chosen path, but also that the packet does not take the unwanted path involving censorship [48]. The proof of taking one path is unnecessarily equivalent to not taking any other path. For example, misbehaving routers can simply process packets without adding required proofs to them [58].

**Destination's demands.** The receiver specifies path preferences for incoming packets mainly for service chaining [58]. Specifically, the receiver may require incoming packets go through services like accounting, inspection, and load balancing. Different service chaining may direct packets through different paths. This requires adapting the Internet to multipath routing [43], which provides two end hosts with multiple path choices instead of a single best path (e.g., the shortest one [70]).

**Intermediate router's demands.** It would seem that intermediate routers have less incentive to accommodate path preferences of end hosts. However, the ability of intermediate routers to satisfy path preferences has a direct impact on the end host's perceived service quality. If service quality is not satisfactory, it will discourage end hosts from continuing their subscription and therefore reduce profits for the service provider. Once intermediate routers cooperate with end hosts, they should collaboratively detect deviated packets as early as possible [46]. Tolerating deviated packets may waste network resources, impair service quality, expose the source's privacy, and jeopardize the destination's security.

### 2.3 Path Validation = Path Enforcement + Path Verification

The path validation problem is introduced to enforce and verify network paths. Kim *et al.* [46] define the goal of path validation as follows.

> **"The Goal of Path Validation [46]:** The source, intermediate routers, and the destination should be able to validate that the packet indeed traversed the path known to (or selected by) the source. Successful path validation ensures that the packet traversed each honest router on the path in the correct order."

To achieve this goal, a path validation solution should feature two functionalities. First, it should enable the source and intermediate routers to enforce mutually agreed upon path preferences. Second, it should also allow the destination and intermediate routers to verify path compliance. The design key is to provide two packet-path bindings. One binding is between a packet and the path it should take. This enables path enforcement, as the binding specifies where to forward the packet. The other binding is between the packet and the path it really takes. This enables path verification as the binding verifies whether the taken path is exactly the specified one. If the verification passes, path validation succeeds.

Path validation solutions use packet-carried states. The fundamental premise of packet-carried states is that bandwidth is plentiful [58]. But it is still important to limit state size such that bandwidth consumption is economized. Specifically, in the case of path validation, the states are initialized by the source and are included in expanded packet headers. The states should include the information of specified paths such that intermediate routers can verify whether they pertain to a correct path, and determine where to correctly forward a packet. An intermediate router need also verify whether a packet went through upstream routers in the correct order. To this end, each intermediate router should leave its mark in the states to prove this to both downstream routers and its destination. The source, intermediate routers, and the destination should all use mutually exchanged shared secrets (e.g., keys) to update and verify states.

Various design challenges arise from the requirement that path validation should function properly in an adversarial, decentralized, and high-speed environment [58]. Path validation may be employed by all three types of network entities (i.e., the source, intermediate routers, and the destination) for purposes of improved service quality, security, and privacy. It is certainly wise to presume the existence of external attackers or even internal misbehaving entities. Such adversaries aim to allow packets that have not taken the specified path to still pass path verification. Path verification should be decentralized so that intermediate routers and the destination can locally verify path compliance. By not transiting packets and their paths' metadata to a centralized server, decentralized path verification avoids corresponding transition delay. Furthermore, it avoids the single point of failure that the centralized server may otherwise encounter. However, limiting the delay induced by processing packet states beyond high-speed forwarding is still a major challenge. For Internet companies, a few milliseconds of delay would turn to a profit decrease of millions of dollars [14].

Clearly, routers need to be upgraded to support path validation using packet-carried states with the above challenges addressed. The upgrade is necessary for both hardware (e.g., for cryptographic computation) and software (e.g., for processing states). This is why path validation solutions still have not been deployed, and remain transcendental for future secure Internet architectures [11]. One question to ask is whether we really have to change router architectures for path validation. If the answer is yes, to what extent do we need to change the current Internet? Researchers on path validation, of course, are not the first to concern themselves with uncontrollable forwarding. How did other peer researchers tackle this problem? Can they achieve path validation, even without modifying Internet protocols? Before discussing path validation solutions in Section 4, let us first review various related solutions and analyze where they fall short for validating network paths in Section 3.

## 3 THE ROAD TO PATH VALIDATION

In this section, we review seemingly qualified solutions and examine where they fall short of path validation. While most of them claim to achieve similar goals as that of path validation, they cannot achieve both path enforcement and path verification, which is the key to path validation. The unqualified solutions far outnumber the qualified ones to be reviewed in Section 4, making the classification process nebulous and challenging. This, however, further inspires us to conduct this thorough and systematic survey of what works for path validation, what does not, and what is worth pursuing.

- We start with secure routing, which secures the process of finding feasible paths between end hosts (Section 3.1.1). Secure routing can be easily confused with path validation as its solutions usually claim secure end-to-end communication as the goal. However, routing plays only a partial role in the communication process. It is forwarding that decides whether packets are delivered in accordance with routing policies. Forwarding decisions are congruent with routing policies only when there exist no faulty or malicious routers.
- To gain more control over the forwarding process, non-security–based solutions like source routing (Section 3.1.2) and traceroute (Section 3.1.3) augment packet delivery with operations for enforcing and verifying paths. Specifically, source routing adds a path directive in packet headers to direct enroute routers on where to forward packets. On the other hand, traceroute requires routers to log packet histories or mark packet headers with visiting histories. Such historical information enables end hosts to reconstruct the path taken by packets and therefore evaluate the compliance with forwarding requirements. However, these solutions focus more on effectiveness rather than security. Both path directives and packet/router histories are not cryptographically secured. This renders them vulnerable to forging and tampering.
- Cryptographic enhancements are then introduced toward secure traceroute (Section 3.2) and secure source routing (Section 3.3), respectively. Secure traceroute can achieve path verification while secure source routing can achieve path enforcement. However, they are not designed to operate concurrently. In other words, it is infeasible or inefficient to simply integrate them for path validation.

### 3.1 What's (Definitely) Not Path Validation

We start with non-security–based solutions that 1) secure routing instead of forwarding (Section 3.1.1), 2) improve forwarding flexibility and robustness instead of security (Section 3.1.2), and 3) support forwarding diagnosis with tamperable traceroute (Section 3.1.3).

*3.1.1 Secure Routing.* For researchers new to path validation, they might easily confuse it with secure routing. Secure routing protocols aim to protect the path finding process against various attacks like prefix hijacking, route spoofing, and eavesdropping [17, 38]. In other words, the ultimate goal of secure routing is to guarantee that any computed route is the best and authentic choice generated by the routing protocol in use [50]. This may elicit an illusion that secure routing promises secure end-to-end communication. However, the caveat is that routing refers to only the process of finding routes; whether packets are forwarded along their associated routes are beyond the scope of routing. Secure routing protocols alone are not sufficient. Via a compromised router, an attacker can still drop, tamper, or arbitrarily forward packets. Therefore, securing the forwarding process (e.g., by path validation) is crucial for secure communication [78].

To re-enforce the difference between secure routing and path validation, we review typical solutions that represent the evolution of secure routing. S-BGP [44] stands out among the first wave of routing security solutions. It adopts cryptography to authenticate destination addresses and route announcements. Specifically, a Public Key Infrastructure (PKI) couples each AS with a public/private key pair. The private key is for an AS to sign its announcements and the public key is for other ASes to authenticate signed announcements. The PKI generates an AS's keys based on its IP address. This way, an AS cannot announce a false IP address and prevents prefix hijacking. Moreover, since an AS does not know the private keys of other ASes, it cannot tamper with route announcements. Follow-up solutions strive to improve S-BGP's efficiency. For example, Secure Path Vector (SPV) [39] adopts the HORS signature [67], which can be computed completely by symmetric cryptography primitives, instead of asymmetric (as required by S-BGP). The evaluation results of SPV demonstrate a 22-time speedup over S-BGP. To alleviate the overhead for a centralized PKI, SCION [88] divides the entire network into a number of *trust domains* comprising core ASes and normal ASes. Each trust domain locally maintains path query service. Communication between different trust domains relies on core ASes. Most typical secure routing solutions targeting the current Internet emerged before 2010. Subsequent research focus has shifted to evaluating their deployment feasibility and practical effectiveness [50]. We refer interested readers to comprehensive surveys on secure routing for more specifics [3, 17, 38, 40, 50, 59]. We emphasize to the reader that constructing paths during routing and enforcing those paths during forwarding are two very different tasks.

*3.1.2 Sourcing Routing.* Source routing embeds the information of source-chosen paths (i.e., path directive) in packet headers to direct routing decisions on enroute routers [35]. The idea was originally adopted by the Internet Protocol but not widely used [6]. It has regained the attention of researchers as more control over packet delivery has been considered necessary [6, 35, 78, 83]. Source routing solutions focus mainly on robustness against link failures and flexibility for policy enforcement. However, they do not consider security to be a design goal. In other words, source routing assumes that routers are benign and faithfully follow path directives to forward packets. Consequently, path directives appear in plaintext. Once routers are compromised, they may violate source routing (e.g., by tampering path directives) without detection. Path validation, on the other hand, considers compromised routers as a possible threat to forwarding, and takes precautions against them.

Although being insufficient for path validation, source routing can greatly improve forwarding flexibility and robustness. It may be used to ease policy enforcement [34, 62, 72, 77, 83, 86]. For example, Internet Indirection Infrastructure (i3) [72] enables a source to specify a chain of services in a packet header. Specifically, each destination providing a certain service is assigned a unique ID. A source specifies a chain of IDs in a packet header and sends the packet to a server. The server then pushes the packet to destinations upon receiving their requests. Similarly, Delegation-Oriented Architecture (DOA) [77], NUTSS [34], and Dysco [86] specify the sequence of middleboxes a packet needs to traverse. NIRA [83] reduces the size of path encoding, while Rule-Based Forwarding (RBF) [62] lets destinations specify the path for packets from a certain source. Source routing can also be applied to multipath routing for a more robust solution [55]. Multipath routing broadens path selection from a single best path, by allowing multiple paths between end hosts. The forwarding process can quickly switch to another path once a link failure occurs, without suffering from the delay caused by path re-computation [47]. Typical multipath routing solutions [31, 35, 55, 84, 90] use source routing in a similar way but differ from each other in how paths between end hosts are found. For example, one may simply calculate two disjoint paths between endpoints [90]. One may also specify multiple next hops for each destination on each intermediate router [55]. Similar to routing deflection [84], the source inserts indicator bits in the packet header to direct each hop which next hop to forward the packet to. To reduce the size of routing tables, Pathlet [31] proposes dividing a path into fragments called pathlets. Each router then maintains routing tables with terminal routers of its residing pathlets as destinations. The source inserts IDs of the chosen pathlets in the packet header to specify the entire path.

*3.1.3 Traceroute.* Traceroute (also known as traceback) is another technique that may easily be confused with path validation. The key idea of traceroute is to enable routers to track traffic history, either by directly marking packet headers [68], or by locally logging packet metadata [69]. Then the destination/receiver can use in-packet router marks, or query packet metadata from one previous-hop router after another, to find out the path taken by received packets. Traceroute was originally proposed to detect IP spoofing and related DDoS attacks [69, 85]. It seems that we can also employ traceroute to assist in path validation. For example, we first let the destination know the specified path for a packet. After receiving the packet, the destination can obtain its taken path via traceroute. The destination then compares the taken path with the specified path. If they match, the packet may have followed the specified path; otherwise, the forwarding process deviates from source-destination agreement. Such a naive application of traceroute, however, fails to meet the requirements for path validation defined in Section 2.3. First, traceroute does not regulate which path a packet should take. Therefore, it does not embed any path directive in packet headers; this fails the requirement of path enforcement. Second, both in-packet marks and on-router logs are not cryptographically secured, making them vulnerable to forging and tampering—particularly if some routers are compromised. Path validation, on the other hand, can effectively combat the impact of compromised routers on packet forwarding.

Most traceroute solutions use packet marking, packet logging, or both. All such methods require modifications to router software. Packet marking requires routers to mark their information (e.g., identifiers) in the headers of passing packets. Due to the large volume of network traffic, Probabilistic Packet Marking (PPM) proposes that each router marks a packet with a pre-defined probability [68]. It is highly likely that each packet carries the information of only some enroute routers. Therefore, a sufficient number of marked packets are needed to reconstruct the path between the source and the destination. Follow-up work [33, 71, 81] strives for improving efficiency, particularly in reducing the number of marked packets. At the other extreme, Deterministic Packet Marking (DPM) requires each router to mark every packet [22, 76]. A single marked packet would suffice for retrieving its taken path. DPM controls packet-header size by compressing router marks in a Bloom Filter [15]. Without touching packets, packet logging enables each router to locally store a digest of every passing packet [69]. The destination can use a single query packet to retrieve packet digests on corresponding routers and reconstruct paths accordingly. Again, Bloom Filters are used to improve storage efficiency. Leveraging the centralized management promised by SDN, NetSight moves packet logs from forwarding devices to centralized servers [37]. It greatly reduces storage overhead on forwarding devices and eliminates false positives caused by Bloom Filters. Packet logging and packet marking are also jointly adopted in some hybrid designs [25, 51, 52, 82] to reap the benefits of both worlds.

## 3.2 What's One Step Closer: Path Verification

Path verification enables the destination to retrieve paths taken by packets and verify their authenticity [23, 60]. Its essential idea is *secure traceroute*. Different from conventional traceroute solutions (Section 3.1.3), secure traceroute uses cryptography to secure the metadata involved in retrieving packet trajectories. Once the actual paths taken by packets deviate from the expected ones, path verification may also help detect the faulty or compromised routers on which path deviation occurs. This, however, does not mean that all solutions for detecting faulty or compromised routers (e.g., [10, 32]) are capable of path verification. Since path verification solutions do not enforce paths, they fail to achieve path validation.

Typical secure traceroute requires that the source probe enroute routers hop by hop, prompting each router to respond with a MAC-protected address of its next hop [60]. With such responses, the source can reconstruct the path taken by probe packets. However, compromised routers may faithfully respond to probe packets yet maliciously discard or detour production packets. It is then necessary for secure traceroute to make probe packets indistinguishable from production packets. The solution in [60] requires a secret channel for the source to synchronize indicators for probe packets with each router. Then the source embeds such indicators in the headers of probe packets. For computing the MACs of next-hop routers' addresses, the source should establish a key pair with each router. To ease deployment, the solution in [79] removes the use of the secret channel and pre-established key pairs. Specifically, it requires the source and each router to share secrets offline before probing. The source directly embeds such secrets in production packets. Each router locally stores the secrets, and based on them, follows simple match-forward to respond to the source, without heavy cryptographic computation. Clearly, probe-based path verification cannot perform at a per-packet granularity; one response from each router to the source per packet would induce significantly high overhead.

Another line of research achieves path verification by analyzing how an aggregate of packets interacts with routers. This method is originally proposed to detect packet loss and delay. Instead of triggering a response to the source upon arrival of individual probe or production packets, each router responds to the source with statistics like volume and timing after receiving a certain number of packets. For example, AudIt [7] requires that a response contain how many packets from the source are received, from which previous hop they were received and at what time. Based on responses from a series of routers, the source reconstructs the path taken by the aggregate of packets. Statistics in responses also helps the source to detect packet loss and delays. AudIt does not use cryptography to protect the integrity of responses. Instead, it leverages legally binding SLAs among ISP peers. SLAs restrict an ISP from manipulating their peers' packets. To further hinder response manipulation, RPVM [41] varies the distribution of packet count by dividing the transmission time into a number of time slots. Specifically, assume that sending out $N$ packets takes the source $T$ units of time. Then source first divides $T$ into $K$ time slots, each with duration of $t_0, t_1...t_i...t_{K-1}$ where we have $\sum_{0 \leq i \leq K-1} t_i = T$. By varying the length of each duration, the number $n_i$ of packets in the slot with duration $t_i$ may have a different number of packets to send. The source timestamps each packet upon sending it out; each router uses timestamps to cluster packets belonging to the same time slot. Then each router should send its monitored distribution of packet count among time slots to the source. Using the received statistics, the source reconstructs the path taken by the packets and notices to which router packet loss or delay might have occurred. Although promising higher efficiency than the probe-based solutions, the lack of cryptographic enhancements renders the statistics-based solutions vulnerable to attacks. For example, we observe a *replacement attack* for malicious routers to drop packets yet evade detection. To launch a replacement attack, a malicious router simply drops some packets and injects an equal number of attack packets.

To attempt unforgeable packet-level history, Secure Packet Provenance (SPP) [23] maintains a cryptographically secured history per packet on each router. SPP is inspired by a network forensic tool called Secure Network Provenance (SNP) [89], which securely records network events and causal links among them. In SPP, packet related events are represented by which router sends which packet via which port at what time. The log of packet events is tamper-evident using the PeerReview log system [36]. The source, the destination, or any intermediate router can query packet histories from a certain router. The routers that respond to the query constitute the path taken by the packet. Their order can be determined via the timings in the log.

### 3.3 What's Almost There: Path Enforcement

We end this section with path enforcement solutions that further pave the road to path validation. Path enforcement is built on top of *secure source routing*, which is easily confused with secure routing (Section 3.1.1) and source routing (Section 3.1.2). Secure source routing differs from secure routing in that the former protects the forwarding process while the latter protects the routing process. It differs from source routing in that it focuses more on security and makes in-packet path directives hard to forge or tamper. Clearly, path enforcement requires cryptography to secure path directives carried in packet headers [9]. It, however, does not require enroute routers to embed verifiable proofs in packet headers. Upon receiving a packet, downstream routers and the destination have no means to verify the actual path taken by the packet. This lack of path verification renders path enforcement alone insufficient for path validation.

Path enforcement needs to prevent forwarding policies specified by in-packet path directives from circumvention. To this end, Platypus [63] introduces the concept and implementation of *network capabilities*. One of the key components in a capability is a waypoint, which specifies a must-visit transit point required by the source. When all routers on the specified path are designated as waypoints, Platypus promises path enforcement. Achieving this goal requires that capabilities be cryptographically hardened against tampering. Platypus introduces *bindings* to ensure capability validity. Specifically, a binding is a cryptographic function of a capability, packet content, and a secret key held by the source of the packet [63]. Waypoints use bindings to verify whether the source has the correct credential for requesting the given service specified in a capability. In other words, routers use capabilities to direct packets, and use bindings to verify the integrity of capacities. This ensures that packets are forwarded along their designated paths. Avramopoulos *et al.* [9] improve robustness by introducing *receipt acknowledgements* and *fault announcements*. Receipt acknowledgements enable each intermediate router and the destination to signal the receipt of a packet, both to upstream routers and to the source. Fault announcements enable every intermediate router to alert both upstream routers and the source when it fails to receive acknowledgements from its downstream router. Once the source receives a fault announcement, it chooses a different path. Ethane [21] applies the idea of path enforcement to enforcing traffic management policies in an intranet. Advertised Reliable Routing over Waypoints (ARROW) [61] further advances incremental deployability, high availability, and high robustness. Taking communication anonymity into account, HORNET [24] integrates the ideas of Platypus [63] and Onion Routing [74, 75]. It encrypts not only path metatdata in packet headers but also packet payloads. Each router can extract from an encrypted packet only its previous hop and next hop. This achieves path enforcement yet reveals no more path information than what is necessary for correct forwarding.

## 4 FINALLY, PATH VALIDATION: PATH ENFORCEMENT + PATH VERIFICATION

In this section, we review path validation protocols that not only enforce source/destination-chosen paths but also verify whether the paths are really taken by packets. To our surprise, after excluding the inapplicable solutions in Section 3, there are only a limited number of path validation solutions. We suspect that this rarity might be mainly due to a common confusion between routing and forwarding. As previously mentioned, most solutions for protecting the routing process claim that they can promise secure end-to-end communication. Such a claim holds upon an implicit assumption that once the best choice of path selection is generated, network entities will faithfully direct corresponding packets along the path. Without a vigilance for the difference between routing and forwarding, readers may conceive secure packet delivery as a fully investigated topic and thus few of them tend to work on it. Path validation, however, is an indispensable building block for future secure Internet architectures. Our survey helps clarify the necessity of path validation, provide researchers with a comprehensive understanding of current solutions, and explore future research directions.

Path validation protocols build themselves on a fundamental premise that network bandwidth is plentiful—incorporating required mechanisms in packet headers is more important than optimizing so caused bandwidth consumption [20]. Their major goal is to jointly guarantee path enforcement and verification. Path enforcement and path verification are achieved via embedding in packet headers with path encodings and router proofs, respectively. These embedded data should be hard to forge or tamper. Their size can be rather large, say hundreds of bytes.

- Since the first framework of path validation [20] (Section 4.1), subsequent solutions strive for reducing the data size to embed and accelerate the computation over them [18, 19, 46, 58] (Section 4.2-Section 4.4). We, however, find that the efficiency enhancements come at the expense of security degradation.
- Although the design framework for path validation is set, the computation method is a promising design space. For example, fast algebraic computation can sometimes be equally effective and secure yet significantly more efficient than expensive cryptographic computation [18, 19] (Section 4.4).

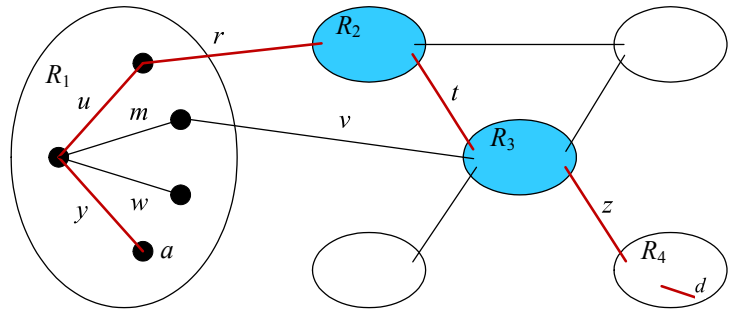

Fig. 1. PFRI instance by rendering Figure 3 in [20]. End-host $\alpha$ in realm $R_1$ sends a packet to end-channel $d$ in realm $R_4$ with realms $R_2$ and $R_3$ providing transit service. The initial partial path is $\alpha yurtzd$. Partial paths connecting $rt$, $tz$, and $zd$ will be locally decided in realms $R_2$, $R_3$, and $R_4$, respectively.

- Formal security proofs for path validation protocols do not keep up with the pace of protocol evolution. So far, only the protocols presented in [46] are validated by formal methods [87]. Since path validation is indispensable for future secure Internet architectures, formal security proofs would be critical for deployment of any solutions (Section 6.1).
- Current path validation protocols still limit in various aspects, such as security, privacy, and efficiency. We in this section focus mainly on their principles and defer discussing their limitations that drive future research directions to Section 6.

### 4.1 Rationale: PFRI

Postmodern Forwarding and Routing Infrastructure (PFRI) [20] is the first work to advocate that enforcement and verification be jointly performed on network paths. It centers on a datagram delivery service, pertaining to the larger clean-slate Postmodern Internet Architecture (PIA) [12]. In subsequent literature on path validation, only ICING [58], to our knowledge, identifies PFRI as a predecessor. Most of the related work, however, refers to ICING as the first path validation protocol. This is likely because PFRI provides only a relatively high level framework of how to design and bootstrap a new network model, with many notable advances other than path validation. For example, it adopts location-independent non-hierarchical identifiers to represent end-hosts. This resolves the mobility constraint of location-dependent IP addressing. It further models a network as a collection of *realms*. A realm consists of nodes (i.e., end-hosts and forwarding elements) and channels (i.e., links between nodes). A sequence of channels from source to destination constitutes a PFRI path. Successful forwarding along a PFRI path necessitates certain centralization [58] to maintain, for example, mappings from identifiers and channels to realms.

PFRI validates the path traversed by a packet by introducing four additional fields over packet headers—*motivation*, *accountability*, *knobs*, and *dials* [20]. The fields of motivation and knobs enforce network paths while the fields of accountability and dials verify them. For path enforcement, the motivation field carries a network path in the packet header as a forwarding directive. For example, in Figure 1, node $\alpha$ in realm $R_1$ has a packet destined for end-channel $d$. Node $\alpha$ first queries its co-realm centralized service for the border channel (i.e., $z$) of the realm (i.e., $R_4$) containing $d$. Then $\alpha$ queries a higher-level centralized service for the transit realms (i.e., $R_2$ and $R_3$) and corresponding inter-realm path (i.e., $rtz$). Using local knowledge of intra-realm topology, $\alpha$ selects an interior path (i.e., $yu$) leading to the border channel $r$ that connects the first transit realm $R_2$. So far, the initial partial path is constructed as $\alpha yurtzd$; it is partial in that the other partial paths connecting $rt$, $tz$, and $zd$ will be locally decided in realms $R_2$, $R_3$, and $R_4$, respectively. Along with the forwarding directive, the motivation field should also carry credentials to convince enroute nodes to forward the packet [20]. Operations other than forwarding can be declared through the knobs field, where credentials are necessary as well. To verify that the path really is forwarded along the path specified in the motivation field, each enroute node should provide a hard-to-forge evidence of packet forwarding in the accountability field [20]. The dials field enriches the types of packet handling to verify. For example, signed nonrepudiable timestamps on enroute nodes help reason about delay.

How to configure the preceding four fields toward path validation, however, is not elaborated in [20]. For example, how does the source compute credentials for enroute nodes? How do enroute nodes compute signatures for the destination? How do all nodes on the path share keys? All these questions may elicit practical concerns and design challenges, especially when mingled with how to control the expansion of packet headers [58].

## 4.2 ICING

Pertaining to another future Internet architecture called NEBULA [5], ICING [58] is acknowledged by most literature [18, 19, 46] as the first concrete design and implementation of path validation. It reinforces the requirements for designing path validation protocols, that is, a path validation protocol should function properly in an adversarial, decentralized, and high-speed environment (Section 2.3). Two key design choices that make ICING efficient are aggregate message authentication codes (MACs) [42] and self-certifying names [4, 54].

First, aggregate MACs reduce state length from quadratic to linear with respect to path length. Consider when ICING enforces a path $P$ that consists of $n$ nodes as $P = (N_0, N_1, ..., N_i, ..., N_{n-1})$, including sender $N_0$, receiver $N_{n-1}$, and routers $N_i$ for $1 \leq i \leq n-2$. For path verification, each $N_i$ for $1 \leq i \leq n-1$ should verify whether a packet with $P$ as its purported path has really been processed by all its upstream nodes $N_j$ for $0 \leq j < i$ in the correct order. To this end, each $N_i$ for $0 \leq i \leq n-2$ should integrate into the packet a number $(n-1) - (i+1) + 1 = n - i - 1$ of proofs, one for each of its downstream nodes $N_k$ for $i < k \leq n-1$. ICING refers to such proofs as *Proofs of Provenance (PoPs)*, which are essentially MACs of the packet. Simply concatenating all PoPs for each $N_i$'s ($0 \leq i \leq n-2$) downstream nodes would make the number of PoPs as large as $\sum_{i=0}^{n-2}(n-2) - i + 1 = \frac{n(n-1)}{2} = O(n^2)$. By using aggregate MACs [42], ICING uses only $n-1$ fields for the preceding path $P$. Each field corresponds to the aggregate MACs/PoPs for $N_i$ ($1 \leq i \leq n-1$) to verify. Specifically, for all $N_i$'s upstream nodes $N_j$ ($0 \leq j \leq i-1$), $N_j$ XORs the MAC it computes to the PoP field for $N_i$. This way, the number of PoP fields shrinks to $n - 1 = O(n)$.

Second, ICING configures shared keys for each pair of nodes to compute/verify PoPs directly using the self-certifying names of nodes [4, 54], without a central naming authority or PKI. To generate non-forgeable PoPs, each pair of nodes should share one symmetric key of their own. Exchanging all required pair-wise shared keys for $n$ nodes would incur an $O(n^2)$ complexity when using PKI. Adopting self-certifying names [4, 54], however, can remove the need of PKI and enable a node to compute its shared key with another node on the fly. Specifically, a node locally computes a public/private key pair. It then locally stores the private key and announces the public key to other nodes; the public key will also function as the name of the node. All keys should be generated in such a way that for any pair of nodes $N_i$ and $N_j$, when $N_i$ computes a key $k_i$ out of its private key and $N_j$'s public key while $N_j$ computes another key $k_j$ out of its private key and $N_i$'s public key, $k_i$ and $k_j$ should be identical. Then $k_i$ and $k_j$ can be used by $N_i$ and $N_j$ as their shared key, respectively. For $N_i$ to prove to its downstream $N_j$ that $N_i$ has processed a packet, $N_i$ computes $MAC_i$ of the packet using key $k_i$ and adds $MAC_i$ in the packet header. Upon receiving the packet, $N_j$ computes $MAC_j$ of the packet using key $k_j$ and then compares $MAC_j$ with $MAC_i$. If $MAC_j = MAC_i$, $N_j$ successfully verifies the history of $N_i$ processing the packet.

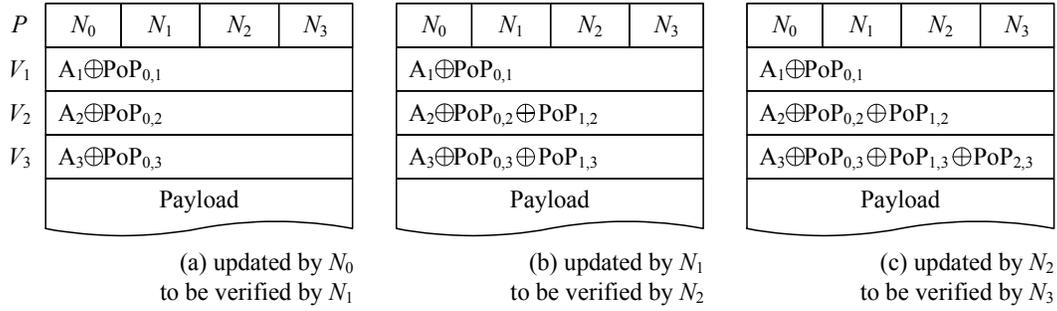

Fig. 2. ICING packet instance by rendering Figure 2 in [58].

Figure 2 illustrates how ICING enforces path validation over a packet along path $P = (N_0, N_1, N_2, N_3)$. $N_0$ and $N_3$ are the sender and receiver, respectively, while $N_1$ and $N_2$ are routers. How $N_0$ finds path $P$ is the focus of routing, which is orthogonal to the forwarding process concerned with ICING [58]. Prior to sending packets along path $P$, $N_0$ should first ask for consents from all the other nodes. ICING refers to such consent as *Proofs of Consents (PoCs)*. $N_i$ ($1 \leq i \leq 3$) computes $PoC_i$ using the node list in $P$ and its private key. After obtaining all PoCs, sender $N_0$ initializes the packet as follows. First, it adds the node list of $P$ in the packet header. For proving to $N_i \in P$ ($1 \leq i \leq 3$) that all nodes along path $P$ give consent, $N_0$ should also include obtained PoCs in the packet header. To prevent an attacker from eavesdropping PoCs, $N_0$ further obfuscates them using the content of the packet. The obfuscation results are called authenticators $A_i$ ($1 \leq i \leq 3$). Each occupies a verifier field $V_i$ for $N_i$ to verify. What $N_0$ should further provide $N_i$ to verify is $PoP_i$. As illustrated in Figure 2(a), $PoP_i$ is XORed to $V_i$, which is currently equal to $A_i$, using the aggregate MAC technique. $N_0$ then sends the packet to $N_1$. $N_1$ needs to verify only $V_1$. The verification succeeds if two conditions are satisfied. First, $N_1$ has issued a PoC of path $P$ to $N_0$. Second, $N_0$ has added in verifier $V_1$ a PoP of the packet. To verify the two conditions, $N_1$ first computes the PoC and then the corresponding authenticator $A_1$. $N_1$ then computes the PoP using the shared key with $N_0$. After obtaining the two values, $N_1$ further compares their XOR result with the value of $V_1$. If they are equal, the packet passes verification and is then updated by $N_1$ with its PoPs for $N_2$ and $N_3$ XORed to $V_2$ and $V_3$, respectively (Figure 2(b)). $N_2$ and $N_3$ then similarly verify and update the packet. Only when the packet visits every node on path $P$ in order can it pass verification on every node.

We observe that the security level of ICING is bounded to one verifier instead of seemingly all verifiers. This is because each node $N_i$ on the specified path verifies only its own verifier $V_i$. Since verifier $V_i$ is an XOR result of several same-size messages (i.e., an authenticator $A_i$ and a number $i$ of PoPs from all $N_i$'s upstream nodes), an attacker does not have to forge every message to forge $V_i$. The attacker only needs to forge a message with the same size as that of $V_i$ via, say, random guessing. The size of a verifier per ICING design is 42 bytes [46]. It renders the success probability of a random guessing attack as negligible as $\frac{1}{2^{8 \times 42}} \ll 0.0001$.

## 4.3 OPT

Origin and Path Trace (OPT) [46], integrated in the SCION Internet architecture [11, 88], achieves higher efficiency than ICING does, given a higher level of trust among on-path nodes. For example, when the destination and all routers on an $n$-node path trust the source, OPT can enable them to verify a packet by only $O(1)$ MAC computations instead of the $O(n)$ MAC computations required by ICING. Consider again $P = (N_0, N_1, ..., N_i, ..., N_{n-1})$ as the path to enforce. Since all nodes $N_i$ ($1 \leq i \leq n-1$) trust the source $N_0$, each $N_i$ generates a shared symmetric key $k_i$ with $N_0$ and sends $k_i$ to $N_0$. $N_0$ initializes a Path Validation Field (PVF) as $PVF_0$ in the packet header. Should each downstream $N_i$ update $PVF_{i-1}$ by $k_i$-keyed MAC, source $N_0$ can be aware of all correct $PVF_i$'s if the packet follows path $P$. Then for $N_i$ to verify whether its received packet has visited $N_0$ through $N_{i-1}$ in the correct order, $N_0$ can compute a shared secret with $N_i$ using the correct $PVF_{i-1}$ and $k_i$. OPT refers to such secret as Origin and Path Validation (OPV) and allocates one $OPV_i$ field for each $N_i$ in the packet header. Upon $N_i$ receiving the packet, it employs the same method as $N_0$ to compute $OPV'_i$ using $PVF_{i-1}$ carried in the packet and $k_i$. Path validation succeeds if $OPV_i = OPV'_i$, costing $N_i$ only $O(1)$ MAC computation.

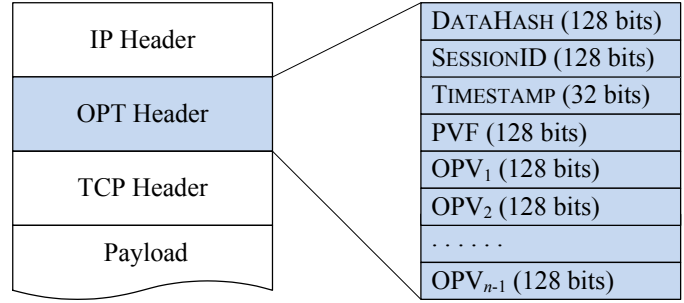

Fig. 3. OPT packet instance by rendering Figure 4 in [46]. Different from ICING, OPT does not include the list of on-path nodes in a packet header. Instead, the source forwards the path information to on-path nodes during they exchange keys. The source $N_0$ precomputes PVF and $OPV_i$ for $1 \leq i \leq n-1$. $N_i$ validates the packet by verifying whether $OPV_i$ can be derived by $PVF_i$. If validation succeeds, $N_i$ needs to update only the PVF field with its proof of processing integrated.

Figure 3 illustrates how OPT augments the packet header for path validation [46]. We focus on the newly added OPT header. Before sending out an OPT packet, the source $N_0$ should initialize all OPT header fields with shared keys with the other on-path nodes. Upon a node $N_i$ receiving the packet, it first verifies whether the packet visits all its upstream nodes in the correct order. This verification takes as input the OPT header fields of DataHash, SessionID, Timestamp, PVF, and $OPV_i$. If the packet passes verification, $N_i$ only needs to update PVF before forwarding the packet to $N_{i+1}$. We sketch the processes of initializing, verifying, and updating an OPT packet as follows.

**Initialization on the source.** The source $N_0$ first initializes DataHash and SessionID of an OPT header.

$$\text{DataHash} = H(Payload),$$
$$\text{SessionID} = H(PK_\sigma || PATH_\sigma || T_\sigma),$$

where $H(Payload)$ is hash of the packet's payload, $PK_\sigma$ is the source's public key together with which there is a corresponding private key $PK_\sigma^{-1}$, $PATH_\sigma = (N_0, N_1, ..., N_i, ..., N_{n-1})$ is the source-selected path for packets to be transmitted in a session indexed by $\sigma$, and $T_\sigma$ denotes the time when $N_0$ initiates session $\sigma$. DataHash can be used to check packet integrity and compute OPV fields. SessionID will be used for on-path nodes to generate shared keys with the source on the fly; these keys $k_i$ are already collected by the source during the key steup phase. Timestamp records the time when the source creates the packet; it prevents an attacker from replaying eavesdropped packets. Based on the preceding three fields, the source $N_0$ further initializes PVF and OSV fields as follows.

$$PVF = PVF_0 = MAC_{k_{n-1}}(\text{DataHash}),$$
$$OPV_i = MAC_{k_i}(PVF_{i-1}||\text{DataHash}||N_{i-1}||\text{Timestamp}),$$

where $PVF_i = MAC_{k_i}(PVF_{i-1})$. To make sure that $PVF_{i-1}$ must come from $N_{i-1}$ to pass verification, the identifier $N_{i-1}$ is also included in the computation of $OPV_i$. After initializing all OPT header fields, the source $N_0$ forwards the packet to the next hop $N_1$.

**Verification on intermediate routers and the destination.** Upon receiving the packet from $N'_{i-1}$, $N_i$ ($1 \leq i \leq n-1$) performs path validation by verifying $OPV_i$. $N_i$ needs to derive its shared key $k_i$ with the source on the fly as the following.

$$k_i = PRF_{S_i}(\text{SessionID}),$$

where PRF is a pseudo-random function keyed with $N_i$'s local secret $S_i$. OPT chooses to generate $k_i$ on the fly as this is faster than retrieving it from local cache or memory [46]. Then $N_i$ computes $OPV'_i$ as the following.

$$OPV'_i = MAC_{k_i}(\text{PVF}_{i-1}||\text{DataHash}||N'_{i-1}||\text{Timestamp}).$$

If the resulting $OPV'_i$ is equal to the $OPV_i$ carried in the packet header, path validation succeeds. Otherwise, path validation fails and $N_i$ may drop the packet.

**Update on intermediate routers.** If the packet passes path validation and $N_i$ is not the destination yet, $N_i$ should update the packet and send it to the next hop. OPT requires $N_i$ to update only the PVF as $\text{PVF}_i = MAC_{k_i}(\text{PVF}_{i-1})$. Different from ICING, OPT does not embed the source-selected path in the packet header. Instead, the source sends $PATH_\sigma$ for session $\sigma$ to on-path nodes during shared-key initialization. Then each on-path node locally follows $PATH_\sigma$ to forward packets.

OPT further suggests how to maintain the $O(1)$ MAC computation complexity when nodes do not trust the source. A distrusting source may collude with router $N_i$ to bypass all the routers in between. Consider, for example, when the source $N_0$ colludes with router $N_2$ to bypass $N_1$. Without this bypass misbehavior being detected by $N_i$ ($3 \leq i \leq n-1$), $N_0$ and $N_2$ need manage to make PVF still look as it would have been updated by the bypassed $N_1$. This is easy for $N_0$ because the expected PVF for $N_1$ is $\text{PVF}_1 = MAC_{k_1}(\text{PVF}_0)$ and both inputs of $k_1$ and $\text{PVF}_0$ are known to $N_0$. The countermeasure against a distrusting source by OPT is introducing another $\text{PVF}^D$ for the destination $N_{n-1}$ to verify [46]. To support $\text{PVF}^D$ verification, each router $N_i$ should create a shared key $k_i^D$ with the destination. $N_i$ then uses $k_i^D$ to update $\text{PVF}^D$. Since $k_i^D$ is unknown to the source $N_0$, $N_0$ can hardly forge $\text{PVF}^D$. Under the enhanced OPT, each on-path node performs two MAC computation over two PVFs.

From the enhanced OPT, we observe a subtle but key difference between OPT and ICING that contributes to their efficiency gap. As previously mentioned, when the destination distrusts the source, it creates one shared key with each of its upstream nodes. Then for the destination to verify whether all the upstream nodes did really process the packet, each upstream node should perform one more MAC computation for generating proof to the destination. Similarly, if each on-path node needs to verify proofs from its upstream nodes without trusting the source, each on-path node has to perform one MAC computation for each of its downstream nodes. This renders the number of MAC computation per node by OPT identical to that by ICING, that is, $O(n)$.

### 4.4 OSV

Orthogonal Sequence Verification (OSV) [18, 19] follows the same design principle as OPT [46] but leverages orthogonal sequences instead of cryptographic computations toward faster path validation. As shown in Figure 4, OSV also relies on a PVF field and an Original Validation (OV) field per on-path node. Each node $N_i$ needs to update PVF in a way known only to itself and the source. The source can easily pre-compute expected $\text{PVF}_i$ upon a series of update by $N_0$ through $N_i$. $N_{i+1}$ can thus use $\text{PVF}_i$ as a proof of whether its received packet has visited $N_0$ through $N_i$ in the correct order. To this end, the source pre-computes $OV_{i+1}$ using a shared function with $N_{i+1}$ and takes $\text{PVF}_i$ as an input. Upon receiving the packet, $N_{i+1}$ first computes an $OV'_{i+1}$ using the PVF (i.e., $\text{PVF}_i$) carried in the packet header. If $OV'_{i+1}$ is equal to $OV_{i+1}$ carried in the packet header, path validation succeeds. The major difference herein from OPT is a much faster computation of PVF and OVs based on orthogonal sequences is adopted. We sketch the OSV framework as follows.

| IP Header | | | |
|---|---|---|---|
| version (8 bits) | header len (8 bits) | unused (4 bits) | credential len (8 bits) |
| user ID (32 bits) ||||
| row index (16 bits) || matrix index (16 bits) ||
| credential $c$ (128 bits) ||||
| PVF (640 bits) ||||
| $OV_1$ (16 bits) || $OV_2$ (16 bits) ||
| . . . . . . ||||
| $OV_{n-2}$ (16 bits) || $OV_{n-1}$ (16 bits) ||
| TCP/UDP Header ||||
| Payload ||||

Fig. 4. OSV packet instance by rendering Figure 4 in [19]. Path validation framework is similar to that of OPT. But OSV adopts orthogonal sequences to yield faster computation of PVF and OV fields.

**Initialization on the source.** The source first generates an $m \times m$ Hadamard matrix $H$ that satisfies $HH^T = mI_m$ [8]. All distinct rows (or columns) of $H$ contain elements of either 1 or $-1$ and are mutually orthogonal in that for the result of inner product of any two rows (or columns) we have $h_i \cdot h_j = 0$. This motivates OSV to use vectors of $H$ as on-path nodes' credentials. Consider, for example, when the source $N_0$ chooses $h_0$ as its credential and forwards $h_1$ to $N_1$ as $N_1$'s credential. Then $N_0$ can simply carry $h_0$ in the packet header. Upon verification, $N_1$ conducts an inner product computation over the carried $h_0$ and its local $h_1$. If the result is 0, $N_1$ can choose to believe that the packet does come from the correct source. OSV makes two necessary design choices to guarantee security.

- First, the dimension of $H$ should be large enough to avoid an attacker from forging valid vectors by random guessing. One desirable property of a Hadamard matrix is that given a fixed dimension, there might be many matrix instantiations. For example, a $32 \times 32$ Hadamard matrix can have up to 13,710,027 possible constructions [45]. But the downside is once the source nails down a specific $m \times m$ matrix, there are up to $m$ vectors in use. Once all these vectors are used as credentials on various packets, the source has to generate a new matrix and forward credentials selected therein to on-path nodes. To avoid overhead, the source generates multiple matrices at the same time, providing more possible vectors to use before the next round of matrix re-initialization. One or more vectors from each matrix may be forwarded to each on-path node as its credentials. The matrix index field in the packet header, as shown in Figure 4, indicates to a node which credential to use for path validation.
- Second, vectors carried in packets should be obfuscated to avoid leaking credentials. For example, instead of simply putting $h_0$ in the packet header, the source $N_0$ generates another random sequence $r_0$ and uses $h_0 + r_0$ as $PVF_0$. Then the source further sets $OV_1 = h_1 \cdot PVF_0 = h_1 \cdot (h_0 + r_0) = h_1 \cdot r_0$. Upon path validation, $N_1$ simply verifies whether $h_1 \cdot PVF_0$ is equal to $OV_1$ as carried in the packet header without revealing $h_0$.

Based on the above two design choices, the source initializes path validation credentials in a packet header as follows. For credential computation, the source $N_0$ first communicates with each on-path node $N_i$ to setup a pair of shared keys $\{k_{N_0}, k_{N_i}\}$. $N_0$ then generates a number of Hadamard matrices. From each $m \times m$ Hadamard matricies, $N_0$ chooses $m_i$ vectors and sends them to $N_i$ as its credentials. The rest $m_0 = m - \sum_{i=1}^{n-1} m_i$ vectors are used as $N_0$'s credentials. $N_0$ will use one of these credentials after another for each of the packets to be sent. Consider, for example, $N_0$ uses $h_{N_0}^j$ for the current packet. Then row index in the packet header is set as $j$ while the matrix index is set according to from which matrix $h_{N_0}^j$ is selected. Credential $c$ and PVF are initialized in the same way as the following.

$$c = \text{PVF} = \text{PVF}_0 = h_{N_0}^j + r_{N_0}^j = h_{N_0}^j + F(k_{N_0}|j),$$

where $F$ is a pseudorandom function for generating the pseudorandom sequence $r_{N_0}^j$. Note that $F$ is shared among all on-path nodes. To initialize $\text{OV}_i$, $N_0$ first computes a series of $\text{PVF}_i$'s ($0 \le i \le n-1$) as the following.

$$\begin{aligned} \text{PVF}_i &= \text{PVF}_{i-1} + h_{N_i}^x + r_{N_i}^j \\ &= \text{PVF}_{i-1} + h_{N_i}^x + F(k_{N_i}|j) \\ &= h_{N_0}^j + r_{N_0}^j + \sum_{i=1}^{i}(h_{N_i}^x + F(k_{N_i}|j)), \end{aligned}$$

where $h_{N_i}^x$ is the credential vector $N_i$ will use to verify the packet. Following the design of OPT, OSV sets $\text{OV}_i$ as follows.

$$\begin{aligned} \text{OV}_i &= h_{N_i}^x \cdot \text{PVF}_{i-1} \\ &= h_{N_i}^x \cdot (r_{N_0}^j + \sum_{i=1}^{i-1} F(k_{N_i}|j)). \end{aligned}$$

The second line holds as the inner product of $h_{N_i}^x$ and any of its mutually orthogonal sequence $h_{N_j}^x$ yields zero. With the preceding fields initialized, the source $N_0$ forwards the packet to $N_1$.

**Verification on intermediate routers and the destination.** Upon $N_i$ verifies the packet, it first authenticates the source and then validates the path taken by the packet so far. Source authentication aims to ensure that the packet does really come from the source that shares keys and credentials for initializing the received packet. Specifically, source authentication verifies the field of credential $c$ and succeeds if given the matrix index, all $N_i$'s credential vectors belonging to this matrix should satisfy the following condition.

$$\forall p \; (1 \le p \le m_i), \; h_{N_i}^p \cdot c = h_{N_i}^p \cdot F(k_{N_0}|j).$$

If source validation succeeds, $N_i$ proceeds with path validation, which succeeds if the following condition is satisfied.

$$\exists p \; (1 \le p \le m_i), \; h_{N_i}^p \cdot \text{PVF} = \text{OV}_i.$$

Both condition verifications leverage the fact that the inner product result of two orthogonal sequences is zero.

**Update on intermediate routers.** Before forwarding the verified packet to $N_{i+1}$, $N_i$ needs to update PVF as the following.

$$\text{PVF} = \text{PVF} + h_{N_i}^p + F(k_{N_i}|j).$$

As with OPT, OSV does not include the path-node list in the packet header. The source-selected path thus needs to know which on-path nodes it exchanges keys. $N_{i+1}$ then follows the preceding verification for source authentication and path validation; if it is not the end of the path, $N_{i+1}$ further updates the packet and forwards it to the next hop $N_{i+2}$.

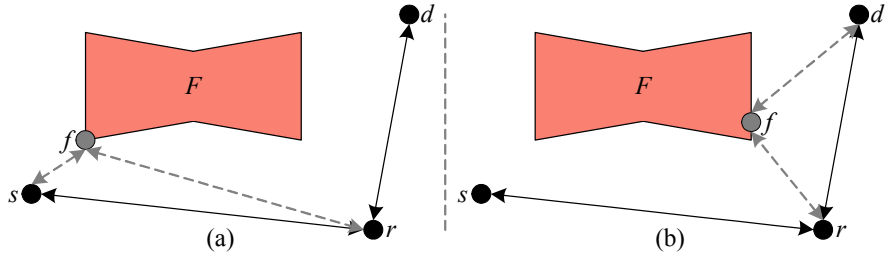

Fig. 5. Alibi Routing (Figure 1 in [48]). For source $s$ to communicate with destination $d$ without packets going through a forbidden region $F$, a relay node $r$ is selected and required to leave proof in the packets. This proof should negate the possibility of the packet traversing the forbidden region $F$. To this end, a relay should be sufficiently far from the forbidden region such that transiting both the forbidden region and the relay induces a noticeable higher delay than transiting only the relay. Let $R(x, y)$ denote the end-to-end latency between two nodes $x$ and $y$. For any node $f$ in the forbidden region $F$, the selected relay $r$ should satisfy that (a) $R(s, r) + R(r, d) \ll \min_f\{R(s, f) + R(f, r)\} + R(r, d)$ if the packet is from $s$ to $d$, and (b) $R(d, r) + R(r, s) \ll \min_f\{R(d, f) + R(f, r)\} + R(r, s)$ if the packet is from $d$ to $s$.

Although computation over orthogonal sequences is faster than cryptographic computation, we observe that the OSV framework resembles OPT under the case of a trusting source. That is, all verifier fields in the packet header can be pre-computed by the source. Other on-path nodes do not share secrets among them. Then for any pair of nodes $N_i$ and $N_j$ ($j > i$) excluding the source, $N_i$ has no means to provide a non-forgeable proof of packet processing to $N_j$. This renders $N_j$ hard to guarantee whether a verified packet has visited all predecessor nodes in the correct order, especially when it does not trust the source.

## 5 WHAT'S BEYOND: ALIBI ROUTING

Path validation solutions surveyed in Section 4 govern where packets should traverse rather than where they should not. It may seem that the two types of governance are identical—Not traversing a node can be achieved by specifying a path without the node included. This holds only when all nodes behave in accordance with the specified forwarding process. But when a certain on-path node misbehaves (Section 6.1), it may direct packets to a forbidden node that is supposed to be avoided. The forbidden node might perform unwanted processing over packets. Then it forwards these packets toward the destination or back to the misbehaving on-path node. If the latter is the case and the forbidden node leaves no mark on packet headers, path validation solutions cannot detect such deviation.

It is, therefore, necessary to design a path validation solution that generates proofs of packets impossibly traversing through some node or region. To the best of our knowledge, Alibi Routing [48], as illustrated in Figure 5, is the only solution for this. The idea is introducing a trusted relay node $r$ that is sufficiently distant from the forbidden region $F$. If a packet traverses through both $r$ and some node $f$ in $F$, it would encounter a noticeably higher latency than when it traverses through only $r$. Consider, for example, when the source $s$ sends a packet to the destination $d$ and expects that the packet avoid the forbidden region $F$. A feasible relay $r$ should satisfy the following constraint.

$$R(s, r) + R(r, d) \ll \min_{f \in F}\{R(s, f) + R(f, r)\} + R(r, d)$$
$$\Leftrightarrow R(s, r) \ll \min_{f \in F}\{R(s, f) + R(f, r)\},$$

where $R(s, r)$ denotes the end-to-end latency between $s$ and $r$; it is estimated using geographic distance between two hosts [48]. Similarly, if the response from $d$ to $s$ still needs to avoid $F$, relay $r$ should also satisfy another constraint.

$$R(d, r) + R(r, s) \ll \min_{f \in F}\{R(d, f) + R(f, r)\} + R(r, s)$$
$$\Leftrightarrow R(d, r) \ll \min_{f \in F}\{R(d, f) + R(f, r)\}.$$

To avoid the forbidden region $F$, the source $s$ needs to collaborate with nodes out of $F$ to find a feasible relay $r$. Source $s$ first locally computes a target region $T$ where a feasible relay $r$ may exist. For computation, $F$ and $T$ are modeled as a set of (latitude, longitude) pairs. Geometrically speaking, $F$ and $T$ are polygons. Locations of $F$'s borders would suffice for computing $T$ based on the preceding constraints. Borders of, for example, countries, can be found online with high precision [1]. The source then sends a query message $< s, d, F, T >$ to one or some of its neighbors. Upon receiving the query message, a node $q$ continues with the search of a feasible relay by satisfying safety and progress conditions. Safety requires that the query message be further forwarded to a trusted next-hop neighbor $n$. That is, the next-hop neighbor $n$ and at least one of its next-hop neighbors should be outside the forbidden region. Given measured RTT—$L(q, n)$—between $q$ and $n$ and estimated minimum possible RTT—$\ell_F(q)$—between $q$ and nodes in forbidden region $F$, a trusted next-hop neighbor should satisfy the following constraint.

$$L(q, n) < \ell_F(q).$$

Several next-hop neighbors of $q$ may satisfy the preceding constraint. Forwarding the query message to which of them should meet the other progress condition. Specifically, if $q$ is not in the target region $T$, the next-hop neighbor $n$ it chooses must be closer to $T$. In contrast, if $q$ is already in $T$, it need also find a next-hop neighbor $n$ that promises a lower end-to-end latency for the source and the destination.

Alibi Routing can guarantee proof of avoidance from the forbidden region if the key used for the relay to generate the proof is kept secret. We, however, observe that Alibi Routing fits better with scenarios where nodes in the forbidden region intend to manipulate packets (e.g., drop or alter them) rather than simply overhear. Because in such scenarios, once the destination receives a packet signed by the relay, it can make sure that the packet is not affected by nodes in the forbidden region. In some other scenarios where nodes in the forbidden region intend to infer certain privacy over passing packets, Alibi Routing becomes vulnerable to attacks by compromised on-path nodes that send copies of packets to the forbidden region, as also concerned in the original Alibi Routing design [48]. Even for encrypted packets, traffic ascription and communication patterns may be inferred [13, 49, 73]. This, however, is hard to prevent and concerned by most path validation solutions. We will discuss more about this concern in Section 6.3, together with other underinvestigated aspects of path validation in Section 6.

## 6 WHAT'S TO UNFOLD: OPEN ISSUES AND FUTURE DIRECTIONS

In this section, we discuss the limitations of current path validation protocols and explore feasible countermeasures. The limitations lie mainly in three aspects: security, privacy, and efficiency.

- **Security**. Since path validation relies on cryptography, it is intrinsically resistant to various security attacks such as forging and replay (Section 6.1). Most path validation protocols (e.g., ICING [58], OSV [18, 19], and Alibi Routing [48]), however, demonstrate their security properties by analysis. So far, only OPT [46] has a formal security proof [87]. We also discuss two other security concerns in the literature. One is how to attest that an on-path node has correctly processed packets as expected (Section 6.2). The other is hidden-node attacks in which some off-path node processes packets without leaving marks on them (Section 6.3).
- **Privacy**. We find that there is a dilemma for path validation solutions to preserve path privacy (Section 6.4). For a node to verify whether a packet has correctly visited upstream hops, it should know the identities of these upstream nodes. On the other hand, for a node to prove its packet processing history to its downstream nodes, it should know the identities of these downstream nodes. Therefore, each on-path node is aware of the other on-path nodes, making it hard to protect path privacy. Furthermore, the leak of path privacy reveals end-host identities and thus prohibits anonymous communication.
- **Efficiency**. Although path validation solutions evolve with faster computation and shorter states, they still cannot efficiently support agile forwarding like multipath routing (Section 6.5). In multipath routing [55, 80], a packet is allowed to take one of multiple paths; it may also switch among allowable paths during forwarding. It becomes prohibitively inefficient to apply current path validation protocols to multipath routing given the likely large number of possible paths. Specifically, dedicating an independent state for encoding each path, concatenating states for all possible paths in a packet header may significantly boost packet length. This may affect the efficiency of both packet processing and packet forwarding.

## 6.1 Formal Security Proof

Security analyses of path validation protocols—ICING [58], OPT [46], and OSV [18, 19]—lack formal proofs and heavily depend on arguing how a path validation protocol defends against common attacks [87]. The major goal of such attacks is typically accompanied with more incentives to fool path validation with packets carrying invalid secrets (e.g., forged or expired), rather than to subvert path validation through, say, simply dropping packets on a compromised router and leaving downstream routers with no packets to verify. We now review typical such attacks and their defeat by path validation protocols.

**Detour.** A detour attack directs a packet either to an off-path router or to an on-path router in an incorrect order. Consider, for example, a source-selected path ABCDE, where A and D are the source and the destination, respectively, and B, C, and D are intermediate routers. In one case of detour, a misbehaving/compromised node, say B, might forward a packet to an off-path node F to evade packet inspection on C. Since the valid path ABCDE should be included in the packet header [58] or informed to each on-path node before hand [46], either case enables the off-path node F to determine the abnormality of receiving a packet that is not supposed to visit it. In the other case of detour, the misbehaving/compromised node B forwards the packet to another on-path node, say D, rather than the correct next-hop C. Whether D does or does not transmit the packet back to C, the secret for D to verify lacks C's proof in it and thus fails path validation.

**Counterfeit.** A counterfeit attack aims to forge valid secrets for detoured or injected packets. When a detour attack skips on-path nodes, it is easily detected due to the lack of proofs from the skipped nodes. A sophisticated attacker would thus need to forge the skipped proofs. This also applies to injected packets that come from the attacker instead of the source. Injected packets essentially resemble packets sent by the source, but altered by the attacker. For both types of packets, the attacker aims to make them pass verification on on-path nodes and finally be accepted by the destination. Since the attacker (i.e., compromised routers) does not have the keys used by other nodes for proof generation, it heavily depends on random guessing to forge secrets. The guessing probability decreases exponentially with proof length and can be negligible when proof length is reasonably large. Counter-intuitively, the attacker need not forge all on-path nodes' proofs to pass path validation. As we discussed in Section 4.2, the weakest link of path validation lies in the destination. That is, if the attacker can successfully forge the proof corresponding to the destination, the attack will succeed whether proofs of upstream nodes can be forged or not. This requires the length of an individual proof field to be sufficiently large. For example, such a proof field by OPT [46] accounts for 128 bits, limiting the probability of random guessing under $\frac{1}{2^{128}} \ll 0.0001$.

**Replay.** Given the difficulty of forging proofs, an attacker may simply record valid packets and replay them later. Such a replay attack may jeopardize both the source and the destination. First, replaying valid packets leads to more than actual traffic attributed to the source. If the packets request certain accountable service from intermediate nodes or the destination, the source will be charged more than supposed. Second, the destination must process replayed valid packets, suffering from a potentially large waste of resources. A high-rate replaying attack thus can be exploited as a DoS attack. All path validation protocols include countermeasures against replay attacks as a key component. An intuitive idea is to let intermediate routers and the destination cache unique metadata from each received packet, such as its carrying proof or counter. Upon receiving a packet, a router or the destination first compares the packet's metadata with the cached data. A match found during the comparison reveals a replay attack. This idea is effective but impractical, due to high storage overhead. A feasible enhancement is to divide the communication between the source and the destination into sessions [46]. Within each session, all packets are embedded with the same timestamp. Received packets with obsolete timestamps (e.g., from days ago) indicate replay attacks and should be discarded. This way, routers and the destination only need to cache packets received within several recent sessions.

**Byzantine/Coward.** A Byzantine attack on path validation is when misbehaving routers violate path validation protocols only when they are less likely to be detected. For example, path validation is invoked when forwarding anomalies such as large delay and packet loss occur, or when the sender or destination has a new service requirement for packets to traverse a path different from the default one. After forwarding anomalies are fixed or the new service is not applied to subsequent traffic, path validation may be switched off from packet forwarding. This makes it hard to detect misbehaviors of a sophisticated attacker (e.g., a compromised router) during the time path validation is not applied. Byzantine attacks are also referred to as coward attacks [46]. In comparison with the preceding attacks, coward attacks raise more challenges but attract less attention.

Both ICING [58] and OSV [18, 19] do not investigate countermeasures. We find this reasonable if path validation is a prerequisite for packet forwarding. That is, every data packet should carry path secrets and it can be accepted only if the secrets pass path validation. In other words, until all on-path nodes generate and exchange necessary keys for verifying and updating path secrets, no data packet is allowed to be transmitted. Note that we use the term "data packet" to differ from the packets for exchanging keys during the setup phase.

An extended version of OPT mitigates coward attacks by removing the apparent key setup process [46]. This way, routers no longer have a clear signal of when path validation takes effect. A malicious node cannot determine when to behave normally per the path validation protocol and when not to without being detected. This forces routers to perform consistently during packet processing and forwarding. The extended OPT assumes that the source and destination trust each other and have shared keys. In contrast, they do not need to exchange keys with intermediate routers before sending packets. Since the source has no shared keys with intermediate routers, it simply initializes $PVF_0$ in the packet header without OSVs. Upon receiving the packet, each intermediate router derives its shared key with the destination on the fly and updates PVF with the derived key. The destination needs to cache received packets for a certain time duration, within which the destination can trigger path validation at any time. Upon a path validation request, each intermediate router reveals the key it uses for updating PVF to the destination.

Although mitigating coward attacks, the extended OPT still requires routers to leave their proofs for future verification. Since packets need to be cached on the destination until verification, they impose a delay. In this sense, we suggest that following the original OPT but allowing no data packet to be sent until successful key setup might be a faster choice, because it allows the destination to instantly verify a received packet. Furthermore, requiring routers to leave proofs—with or without the apparent key setup process—may alert routers of future verification. This renders the problem tackled by OPT as a weak version of the coward attack. A more challenging version with path validation either on or off would be an interesting research direction.

**Collusion.** Two or more compromised/malicious nodes on a path may collude to launch any of the preceding attacks. Consider again a source-specified path ABCDE for example. B and D could collude to skip C. Instead of verifying the packet, D may directly update the packet with its proof and further forward it to E. E will easily detect this detour attack due to lack of C's proof in the packet header. One may wonder what happens when B, C, and D collude. For example, they could skip C but still have C provide its proof. This trickier version of detour attack cannot be detected by E. The intention behind such a collusion aims to make packets evade C's processing but exhibit as if they were processed by C. All path validation protocols consider this as unaddressable. We will discuss it in Section 6.2 and suggest a countermeasure.

Unlike the common and more informal security analysis provided above, only OPT [46] is backed up by formal security proofs [87]. As path validation relates to network protocols, its proof should be applicable to arbitrary network topologies, and therefore raises more challenges than proving cryptographic protocols [87]. Specifically, proofs in [87] focus on the version of OPT with an untrusted source, that is, each router shares keys with both the source and the destination (Section 4.3). Properties to prove include 1) secrecy and authenticity of the key setup phase, and 2) authenticity of origin/source and path of the packet during the forwarding phase. OPT is first formally expressed using $LS^2$ [28] that can reason about programs concurrently executing with programs controlled by adversaries. Adversaries are interface-confined [30] to reasonably limit their capabilities. Moreover, to reason about protocols using cryptographic functions, $LS^2$ is further extended with relevant definitions of data structures and axioms formalized by PCL [27]. Finally, $LS^2$-encoded OPT runs in Coq [2], a security proof assistance system that OPT uses to emulate an adversarial environment. We refer interested readers to [87] for proof details. What we would like to emphasize here is that the proofs of OPT focus on only the chained MAC technique. Proofs over techniques like aggregated MACs by ICING [58] and orthogonal sequences by OSV [18, 19] remain untouched and would be critical for deploying path validation as well.

## 6.2 Packet Processing Attestation

All path validation protocols acknowledge the incapability of verifying whether packet processing takes effect [18, 19, 46, 58]. This concern is reasonable because a router adding its proof on a packet does not guarantee that it faithfully processes the packet as requested. Consider, for example, a router connecting to a middlebox for deep packet inspection. If compromised, the router may not direct packets to the middlebox [16]. Even if the router behaves correctly, the middlebox may still not inspect received packets as expected. In both cases, as long as the router adds its proof to packets and sends them to the correct next hop, downstream routers and the destination cannot detect these uninspected packets using path validation protocols. Attestation of packet processing is imperative for strengthening path validation.

We suggest a probing-based method to indirectly attest packet processing. The idea is to inject probe packets along with production packets. Probe packets are constructed in such a way that the expected result of processing them are known to the destination. Take a Firewall service for example. When the source or the destination buys the Firewall service that filters unwanted traffic, they need to verify that Firewall works correctly and faithfully. To this end, the source can send a probe packet that must be filtered by Firewall. If the destination still receives the probe packet, it detects an inaccurate or likely malicious behavior of Firewall. Knowing this detection method, a sophisticated attacker manipulating the Firewall may normally process probe packets, while still manipulating production packets. Therefore, probe packets should be indistinguishable from production packets. Additionally, probe packets should carry random indicators synchronized between the source and destination. Such indicators help the destination to identify probe packets and discard them to avoid their impact on performance and security.

### 6.3 Hidden-Node Attack

If some nodes on the specified path are compromised, they might detour packets to some off-path nodes, which perform unwanted inspection over the packets without leaving marks, and then direct them back to the specified path. Consider, for example, a specified path $N_0 N_1 N_2 N_3$. Compromised $N_1$ may detour packets to an off-path node $N_x$, which then directs packets back to $N_1$. In another case, if $N_2$ is also compromised and colludes with $N_1$, the off-path node $N_x$ can direct packets back to $N_2$. This protocol breach is referred to as a hidden-node attack [58]. Hidden nodes correspond to the off-path nodes. To avoid being noticed by end hosts, hidden nodes would not interfere with their communication by, for example, altering or dropping traffic. Since hidden nodes do not leave marks on packets, path validation protocols surveyed in Section 4 cannot detect hidden-node attacks. If the hidden node is far away from the specified path and causes a noticeable transmission delay, end hosts can use such a delay to detect hidden-node attacks. In this case, the detection method resembles Alibi Routing [48]. However, if the hidden node is few hops away from compromised on-path nodes, detouring packets through them would not cause too much delay. A slight delay is hard to be differentiated from normal transmission-time fluctuation. This makes the detection of hidden-node attacks very difficult.

A variation of hidden-node attacks do not detour packets through hidden nodes. Instead, compromised on-path nodes simply copy packets and forward the copies to hidden nodes while forwarding the original packets to the next hop on-path node [58]. Alibi Routing suggests that obfuscating packets with anonymity techniques (e.g., Tor [74]) can make packet copies less useful to hidden nodes and thus prevent privacy leakage.

Most path validation protocols consider hidden-node attacks hard to prevent and detect [46, 48, 58]. Their common advice is that the sender and the receiver should choose nodes they trust enough to form the specified path. This way, on-path nodes are less likely to be compromised and to detour packets or send copies to hidden nodes.

### 6.4 Host Anonymity and Path Privacy

Path validation cannot preserve anonymity and path privacy. End-host anonymity aims to protect the identities of the source and the destination during their communication. End-host anonymity implies the requirement of path privacy as well. If the path information is leaked, it is easier to track down end hosts connected by the path. Therefore, anonymous communication protocols vary packet formats per hop to prevent path privacy leakage by packet correlation. The typical protocols for anonymous communication are built upon Onion Routing [75] or its evolved version Tor [74]. In such protocols, the source specifies a series of relay nodes that constitute an overlay network.

After negotiating keys with each relay node, the source encrypts its message $m$ to the destination using nested encryption. Consider, for example, source $S$ and destination $D$ communicate via three relay nodes $R_1$, $R_2$, and $R_3$. Let $k_1$, $k_2$, and $k_3$ represent the keys used for $S$ to communicate with $R_1$, $R_2$, and $R_3$, respectively. Then the encrypted message sent from $S$ to $R_1$ is constructed as the following.

$$\text{Enc}_{k_1}(R_2, \text{Enc}_{k_2}(R_3, \text{Enc}_{k_3}(message, D))).$$

Upon receiving the above encrypted message, $R_1$ decrypts it and can only obtain the specification of the next hop $R_2$ and the encrypted message targeted at $R_2$—$\text{Enc}_{k_2}(R_3, \text{Enc}_{k_3}(message, D))$. Similarly, $R_2$ forwards the encrypted message $\text{Enc}_{k_3}(message, D)$ to $R_3$, which finally forwards the message to $D$. (Note that the message should be encrypted with a key shared between $S$ and $D$ to protect confidentiality.) For each relay node, all it can infer from the received encrypted message is its previous hop and its next hop. Such inferred information is the minimum of any forwarding protocol, and it is maximum revealed by Onion Routing. In other words, besides the first relay node, the other relay nodes (e.g., $R_2$ and $R_3$) cannot know the source. This guarantees the source anonymity. The destination anonymity can be protected in the same way. But path validation protocols require that path information be revealed to on-path nodes through packet-carried states (e.g., ICING [58]) or pre-loaded configuration (e.g., OPT [46]). If this is not the case, a node cannot ensure packet-processing proofs of which previous hops it needs to verify.

### 6.5 Forwarding Agility

Current path validation protocols bind a packet with only one fixed path. Therefore, they cannot efficiently apply to agile forwarding, which allows packets to switch paths during forwarding. Agile forwarding is built on multipath routing. Unlike traditional routing protocols like BGP, multipath routing computes multiple forwarding paths for each packet. There are two types of forwarding. The first type selects one out of the multiple paths and forwards the packet along the selected path [80]. This type of forwarding usually benefits traffic engineering such as congestion control and load balancing. In this case, each packet is still coupled with one path during forwarding; existing path validation protocols can be directly applied. More of our interest is the second type, which allows on-path nodes to switch packets to another backup path [55]. It aims to improve forwarding robustness through, for example, fast failover. A typical such multipath routing, Path Splicing [55], organizes a packet's multiple paths in a tree structure. The number of possible paths thus may be exponential with path length, promising significant forwarding agility. However, it is hard for current path validation protocols to encode such a huge set of paths as packet-carried states. Finding practically efficient path encoding techniques is therefore imperative for path validation under agile forwarding. How to find a tradeoff between forwarding agility and validation efficiency would also be an encouraging research direction.

## 7 CONCLUSION

We have presented the first comprehensive survey of research on validating network paths. Due to various limitations in agility, security, and privacy, traditional network architecture has long been insufficient for fostering innovative Internet services. No matter whether the future Internet follows a clean-slate or evolutionary design [66], validating network paths would be critical for security [11]. Path validation strives for two enhancements over traditional packet delivery. One is path enforcement, which enables routers to forward packets along paths specified by the source, the destination, or intermediate routers. The other is path verification, which enables intermediate routers and the destination to verify whether a packet has taken its specified path.

Both enhancements introduce cryptographically protected states carried in packet headers. In a nutshell, path validation requires that the source specifies in a packet header which routers the packet should traverse in order. Routers then forward the packet according to this path directive. Furthermore, the source and each router should embed in the packet header a series of proofs, one for each downstream node (i.e., downstream routers and the destination). Such proofs allow a node to verify path compliance. All these packet-carried states should be hard to forge or tamper. With path validation, network entities can gain more control over packet delivery. A variant of path validation can even enable end hosts to verify whether their packets detour from a certain region [48].

Although ensuring higher service quality and security, path validation attracts only a small body of work. We find that this rarity might be attributed to a long-standing confusion between routing and forwarding. Routing aims to find paths for pairs of end hosts while forwarding aims to direct packets from one end host to another. Compromised routers may make the forwarding decision deviate from what is generated by the routing protocols. In other words, securing the routing process alone cannot guarantee correct packet forwarding. It is also insufficient to merely enforce packet delivery along a specified path, or to simply verify which path a packet has taken. Path enforcement and path verification should be jointly adopted to accomplish path validation. This strict requirement renders many routing and forwarding solutions unqualified for validating network paths. Clearly, in order to satisfy both path enforcement and verification, the cryptographic states carried in packet headers would be quite large in size. A major trend in path validation design is therefore shortening the states. However, we find that the efficiency gain comes at the cost of security degradation. We conduct a comprehensive analysis of the properties and limitations of existing path validation solutions. It reveals that they are still limited in various aspects, thus opening a wide range of inspiring and exciting research directions to further advance security, privacy, and efficiency.

We hope that this survey article benefits both the research and industry community with useful references and directions for validating network paths. Together, we shall strive for a better Internet.

## ACKNOWLEDGMENTS

The work is supported in part by the National Science Foundation of China under Grant 61402404 and Grant 61602093. The authors would also like to thank Editors and Reviewers of ACM Computing Surveys in advance for their review efforts and helpful feedback.